\documentclass[pra, preprint, showpacs,amsmath,amssymb]{revtex4}

\usepackage{wasysym}
\usepackage{graphicx}
\usepackage{dcolumn}
\usepackage{bm}
\usepackage{xcolor}

\begin{document}

\title{Entanglement and Thouless times from coincidence measurements across disordered media
}

\author{Nicolas Cherroret and Andreas Buchleitner}
\affiliation{
Physikalisches Institut, Albert-Ludwigs-Universit\"{a}t Freiburg, Hermann-Herder-Str. 3, D-79104 Freiburg, Germany
}

\date{\today}

\begin{abstract}
We show that the entanglement of an initially frequency-entangled photon pair propagating through a disordered scattering medium can be read off in transmission from the coincidence counting rate. The very same quantity also encodes the Thouless time of the medium.
\end{abstract}

\pacs{42.25.Dd, 42.50.Lc, 42.65.Lm}

\maketitle
 
\section{Introduction}
 
The study of speckle patterns, irregular interference structures produced by light propagating through an inhomogeneous medium,  is of fundamental importance in the understanding of coherent wave propagation in disordered  media \cite{Berkovitz94}. While, at first sight, a speckle pattern looks perfectly random, closer analysis reveals a more complex picture. In the backscattering direction, for instance, the light intensity is enhanced, due to the constructive interference of pairs of counter-propagating multiple scattering paths. This is known as the coherent backscattering effect \cite{Wolf85}. In transmission, speckle patterns exhibit long- and infinite-range correlations between distant spots. The latter are, in particular, at the origin of universal conductance fluctuations in electron transmission across mesoscopic conductors \cite{Webb85}. Once again, these highly nontrivial correlations originate from the interference between pairs of scattering paths which persists under disorder average, and have been studied extensively during the last decades \cite{Berkovitz94, Akkermans}.

While our knowledge of the statistics of \emph{classical} waves in disordered media is today fairly well developed, little is known about the statistical properties of \emph{nonclassical} light propagating in such media. Recently, an important step forward was made with the investigation of the interplay between quantum fluctuations of light, on the one hand, and the classical fluctuations induced by a scattering medium with static disorder \cite{Lodahl05,Smolka09} or moving scatterers \cite{Skipetrov07}, on the other hand. In this scenario, a given quantum state of light enters a disordered medium, and signatures of the quantum statistics are found in the photon number variance analyzed upon transmission. Later, the probability distribution of the coincidence counting rate, which defines a ``two-photon speckle", was analyzed on output of a disordered medium \cite{Beenakker09}. Two-photon speckles are relevant for studying entangled states of light, and were recently observed experimentally \cite{Peeters10}, but interferences between different scattering paths were assumed to vanish under disorder average.

Interference phenomena with nonclassical light in disordered media have so far only been addressed in the framework of one-dimensional tight-binding models \cite{Silberberg10} or in the transport of quantum states with independent photons \cite{Ott2010}. In our present contribution, we consider the propagation of a pair of \emph{spectrally entangled photons} in an open, disordered medium with three-dimensional disorder. We show that the emerging interference effects bear \emph{all} the information on the spectral entanglement between the two photons, and that, more surprisingly, information on the \emph{dynamical} properties of the disordered medium is also inscribed in the interference signal. Finally, we suggest a method to access this information experimentally. To do so, we analyze the photon coincidence counting rate of a photon pair transmitted through a disordered waveguide, within a continuous-mode approach (Secs. \ref{Model}, \ref{CCR}). We show that interference contributions to the coincidence rate depend on two time scales, a time $t_\eta$ characteristic of the spectral, and, in particular, the entanglement properties of the photon pair, and the Thouless time $t_D$, \emph{i.e.} the typical time it takes for a photon to traverse the disordered sample by diffusion \cite{Thouless74}. In Sec. \ref{ISI}, we examine how the interference contributions depend on these characteristic times. Finally, in Sec. \ref{IDP} we propose a way to measure these two times experimentally, by introducing a small delay between the two photons before they enter the medium. In this situation, we show that interference leads to a peak in the coincidence counting rate, whose size gives direct access to $t_\eta$ and $t_D$.

\section{Model}
\label{Model}

\begin{figure}[h]
\includegraphics[width=12cm]{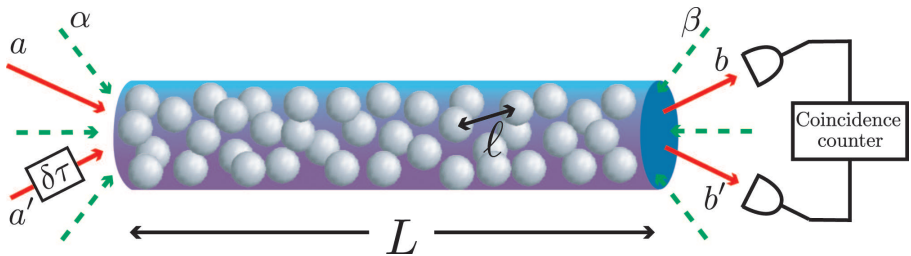}
\caption{\label{model} 
(color online). Two photons, incident in spatial modes $a$ and $a^\prime$, are scattered from a disordered waveguide of length $L$, with a mean free path $\ell\ll L$. The coincidence counting rate between two outgoing modes $b$ and $b^\prime$ is analyzed by photodetectors. The input-output relation (\ref{In-Out}) takes into account the coupling to all unoccupied (vacuum) modes $\alpha$ and $\beta$, which are depicted by dashed arrows.
}  
\end{figure}

Our model is depicted in Fig. \ref{model}. A source produces pairs of photons which are scattered from a disordered waveguide of length $L$ and cross-section $A$. The mean free path, \emph{i.e.}, the average distance between two consecutive scattering processes, is assumed to be small, $\ell\ll L$, such that each photon experiences multiple scattering before being collected on output by two detectors. In the most general situation considered in this paper (Sec. \ref{IDP}), one of the two photons is delayed by a time $\delta\tau$ with respect to the other. 
If we assume perfectly reflecting boundary conditions at the outer surface of the waveguide, only a finite number $2N(\omega)=4\times k(\omega)^2A/(4\pi)$ of transverse spatial modes are supported, where $k(\omega)$ is the wave vector at frequency $\omega$. Note that $N(\omega)$ accounts for two possible propagation directions and two polarization states of the electromagnetic field. Another important quantity characteristic of the disordered medium is the dimensionless conductance $g=(4/3)N\ell/L$. $g$ is the single parameter that controls the transport properties across the disordered waveguide, from the regime of Anderson localization ($g\lesssim 1$) to the regime of diffusive multiple scattering $g> 1$ \cite{Abrahams79}. We will focus on the case $g> 1$ hereafter.

An incoming photon pair is described by a two-photon wave-packet, within a continuous-mode formalism, with the electromagnetic field quantized in the usual way \cite{Loudon}. The two-photon state is written in the general form $|\psi\rangle=\int d\omega d\omega^\prime S(\omega,\omega^\prime)\hat{a}_a^\dagger(\omega)\hat{a}_{a^\prime}^\dagger(\omega^\prime)|\text{vac}\rangle$, where the continuous-mode operator $\hat{a}_\alpha^\dagger(\omega)$ creates a photon with frequency $\omega$ in the spatial mode $\alpha$. The two-photon wave-packet amplitude $S(\omega,\omega^\prime)$ contains all spectral information about the photon pair. In analogy with \cite{Campos90}, we choose
\begin{equation}
S(\omega,\omega^\prime)=\left[\dfrac{1}{2\pi\sqrt{1-\eta^2}/t_c^2}\right]^{1/2}
\exp\left[{\frac{-(\omega-\omega_0)^2-(\omega^\prime-\omega_0)^2+2\eta(\omega-\omega_0)(\omega^\prime-\omega_0)}{4\sqrt{1-\eta^2}/t_c^2}}\right]e^{-i\omega^\prime\delta\tau},
\label{beta}
\end{equation}
which provides a simple and general model of a two-photon wave-packet, characterized by a central photon frequency $\omega_0$, and a single-photon coherence time $t_c$. Experimentally, such a two-photon wave-packet can be produced by means of parametric down conversion, in which two entangled photons originate from a nonlinear crystal irradiated by a ``pump" photon. The dimensionless parameter $\eta\in\left]-1,1\right[$ determines the frequency entanglement of the two down-converted photons, and depends on the coherence time of the pump photon, the length of the crystal, and the phase velocities of the pump and the down-converted photons in the crystal \cite{Grice01}. Different types of entanglement are distinguished by the cases $\eta<0$, $\eta>0$ and $\eta=0$. These correspond to anticorrelated, correlated and uncorrelated photons, respectively. The amplitude $|\eta|$ gives the ``strength'' of entanglement, the limits $\eta\rightarrow \pm1$ corresponding to fully entangled states. Fully anticorrelated photons ($\eta\rightarrow -1$) correspond to $S(\omega,\omega^\prime)\propto\delta(\omega+\omega^\prime-2\omega_0)$ and are produced by standard parametric-down conversion with a monochromatic pump \cite{Mandel}. The possibility to generate fully correlated ($\eta\rightarrow 1$), for which $S(\omega,\omega^\prime)\propto\delta(\omega-\omega^\prime)$, or completely uncorrelated ($\eta=0$) photons, for which $S$ can be written as the product of two independent wave-packets, has also been discussed \cite{Walton03, Grice01}. Recently, an original experimental method was devised to produce photon pairs with arbitrary values of $\eta$, with the help of a superlattice of nonlinear crystals \cite{URen06}. Let us finally note that the Gaussian form of Eq. (\ref{beta}) is chosen for convenience, since the results derived herafter do not qualitatively change for slightly different shapes of $S$.

In order to study the transport properties of the photon pair in the disordered medium, we make use of standard input-output relations between incoming and outgoing annihilation operators \cite{Beenakker98}: 
\begin{equation}
\label{In-Out}
\hat{a}_b(\omega)=\sum_{\alpha=1}^{N(\omega)}t_{\alpha b}(\omega)\hat{a}_\alpha(\omega)+\sum_{\beta=N(\omega)+1}^{2N(\omega)}r_{\beta b}(\omega)\hat{a}_\beta(\omega),
\end{equation}
where $t_{\alpha b}$ and $r_{\beta b}$ are transmission and reflection coefficients from the incoming modes $\alpha$ and $\beta$, respectively, to the outgoing mode $b$. Note that Eq. (\ref{In-Out}) fully accounts for the coupling to all vacuum states $\alpha\ne a,a^\prime$ and $\beta\ne b,b^\prime$ (see Fig. \ref{model}). Creation and annihilation operators obey the usual bosonic commutation relations $[\hat{a}_i(\omega),\hat{a}_j^\dagger(\omega^\prime)]=\delta_{ij}\delta(\omega-\omega^\prime)$.

\section{Coincidence counting rate for $\boldsymbol{\delta\tau=0}$}
\label{CCR}

\begin{figure}[h]
\includegraphics[width=10cm]{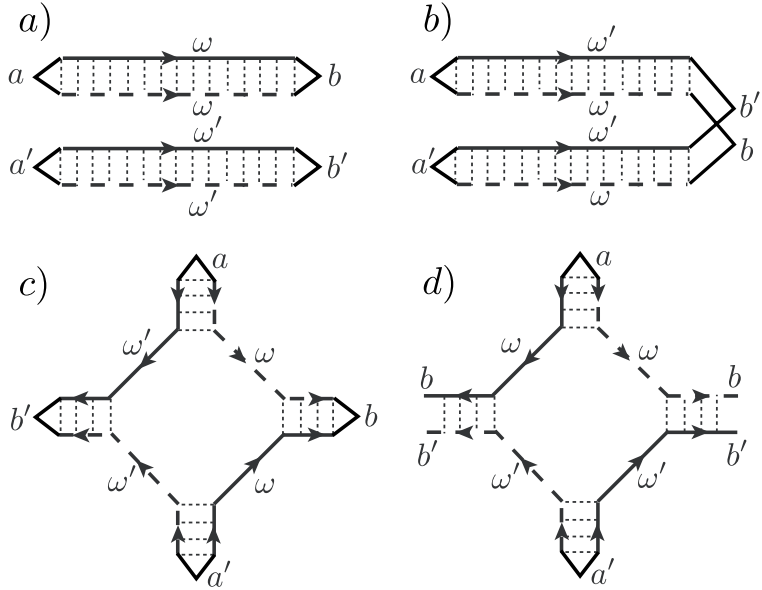}
\caption{\label{Diagrams} 
Leading  order diagrams that contribute to the kernel $K_{aa^\prime bb^\prime}$, and give rise to the four terms in the right-hand side of Eq. (\ref{Main_result}). In each diagram, the two parallel lines connected by dotted ``ladders" symbolize averages of products of transmission coefficients, $\overline{t_{ij}(\omega)t_{i^\prime j^\prime}(\omega^\prime)}$, with incoming $i,i^\prime=a$ or $a^\prime$ and outgoing $j,j^\prime=b$ or $b^\prime$ directions \cite{Footnote2}.
}  
\end{figure}

In this section, we assume that there is no time delay between the incident photons. We first examine the mean photon number $\overline{\langle I_b\rangle}=\int_0^T dt\, \overline{\langle \hat{a}_b^\dagger(t)\hat{a}_b(t)\rangle}$ registered by a photodetector in a given outgoing mode $b$, during a sampling time $T$. This quantity gives the probability to obtain a ``click" from the photodetector, meaning that it collected one photon of the pair. $\langle\cdots\rangle$ refers to quantum mechanical averaging over the two-photon state, and $\overline{\cdots}$ to classical ensemble averaging.  These two types of averaging highlight that the photon number fluctuates due to both, the quantum nature of the light, and the stochastic properties of the disordered medium. The time-dependent operators $\hat{a}_b(t)=(1/\sqrt{2\pi})\int_{\omega_0-\Delta\omega/2}^{\omega_0+\Delta\omega/2}d\omega \hat{a}_b(\omega)\text{exp}(-i\omega t)$ are defined via a Fourier integral over the bandwidth $\Delta\omega\ll\omega_0$ of the photodetector around the central photon frequency $\omega_0 $. $\Delta\omega$ is typically much larger than the spectral bandwidth $1/t_c$ of individual photon wave-packets. In addition, since usual photodetectors are not able to resolve photon wave-packets, the detection time $T$ exceeds any other time scale of the problem, such that we have
\begin{equation}
1/T\ll1/t_c\ll\Delta\omega\ll\omega_0.
\label{Assumptions}
\end{equation}
This separation of time scales allows to extend the range of frequency and time integrations from $-\infty$ to $\infty$, and to approximate the number of transverse spatial modes by $N(\omega)\simeq N(\omega_0)$. We further assume that the properties of the disordered medium do not change at the scale of $\Delta\omega$, such that quantities like the mean free path are constant upon integration. Under these premises, the input-output relations (\ref{In-Out}) and standard diffusion theory \cite{Akkermans} lead to $\overline{\langle I_b\rangle}=(4/N)(\ell/L)$. Note that $\overline{\langle I_b\rangle}$ is independent of $t_c$ and $\eta$, \emph{i.e.} the mean photon number does not contain any spectral information on the two-photon state \cite{Beenakker09}. We therefore turn to the coincidence number $\overline{\langle I_bI_{b^\prime}\rangle}$ between two outgoing modes $b$ and $b^\prime$ (with possibly $b=b^\prime$), which will be the figure of merit from now on. This quantity is defined as
\begin{equation}
\overline{\langle I_bI_{b^\prime}\rangle}=\int_{0}^T dt\int_{0}^T dt^\prime \overline{\langle\hat{a}_b^\dagger(t)\hat{a}_{b^\prime}^\dagger(t^\prime)\hat{a}_{b^\prime}(t^\prime)\hat{a}_b(t)\rangle}.
\end{equation}
After ensemble averaging, this turns into
\begin{equation}
\overline{\langle I_bI_{b^\prime}\rangle}=
\int_{-\infty}^\infty d\omega \int_{-\infty}^\infty d\omega^\prime\left\{
1+2K_{aa^\prime bb^\prime}(\omega,\omega^\prime)\text{Re}\left[S^*(\omega,\omega^\prime)S(\omega^\prime,\omega)\right]
\right\},
\end{equation}
with Eqs. (\ref{In-Out}) and (\ref{Assumptions}). The kernel $K_{aa^\prime bb^\prime}$ accounts for fourth-order interference between photon path amplitudes multiply scattered by the disorder.  These interferences are described by combinations of products of four transmission coefficients, averaged over disorder, and represented by Feynmann diagrams \cite{Akkermans}. In the diffusive limit $g>1$ considered here, $K_{aa^\prime bb^\prime}$ can be expanded in powers of $1/g$. Performing this expansion up to the first order in $1/g$, the coincidence counting rate $R$ reads
\begin{equation}
\label{Main_result}
R=\dfrac{\overline{\langle{I_bI_{b^\prime}}\rangle}}{\overline{\langle I_b\rangle}\times\overline{\langle I_{b^\prime}\rangle}}=
\dfrac{1}{2}\left\{
1+\dfrac{2}{3g}C^{(2)}\left(\dfrac{t_ \eta}{t_D}\right)
+\delta_{bb^\prime}
\left[
C^{(1)}\left(\dfrac{t_ \eta}{t_D}\right)+
\dfrac{2}{3g}C^{(2)}\left(\dfrac{t_ \eta}{t_D}\right)
\right]
\right\},
\end{equation}
which is the main result of the present contribution. In Eq. (\ref{Main_result}) we introduce two characteristic times, $t_\eta=t_c/\sqrt{1-\eta}$ and the Thouless time $t_D=L^2/(\pi^2D)$, with $D$ being the photon diffusion coefficient. Whereas $t_\eta$ contains the spectral information on the photon pair (through the coherence time $t_c$ and the entanglement parameter $\eta$), $t_D$ is a quantity characteristic of the disordered medium, giving the typical time it takes for a photon to diffuse through the sample. The Thouless time was initially introduced by Thouless in the context of electronic conduction in metals \cite{Thouless74, Footnote}, and is, in essence, a quantity characterizing the \emph{dynamics} of the disordered medium, due to its proportionality to the inverse of the diffusion coefficient \cite{Storzer06, Skipetrov06, Albada91}. The dependence of $R$ on $t_\eta$ and $t_D$ is discussed in detail in the next section. Here we first comment on the physical content of Eq. (\ref{Main_result}): its right-hand side contains four terms. The first, constant term describes the independent propagation of the two photons in the disordered medium, and corresponds to the diagram in Fig. \ref{Diagrams}a. The term denoted $C^{(1)}$ is depicted in Fig. \ref{Diagrams}b. It results from a fourth-order interference, said to be \emph{local}, because the exchange of photon path amplitudes (the ``crossing") occurs close to the boundary of the medium. The two terms $\propto C^{(2)}$ and $\propto\delta_{bb^\prime}C^{(2)}$ in Eq. (\ref{Main_result}) encode another type of fourth-order interference, in which photon path amplitudes change partners at some point in the multiple scattering sequence. They are represented by the diagrams of Fig. \ref{Diagrams}c and \ref{Diagrams}d, respectively \cite{Footnote2}. Note that here the photon path amplitudes recombine and then propagate again after the crossing, such that the underlying interference process is \emph{nonlocal}. 

We now discuss the structure of Eq. (\ref{Main_result}). When the coincidence counting rate is measured in one outgoing mode, $b=b^\prime$, such that $R\simeq1/2[1+C^{(1)}+(2/3g)C^{(2)}]\simeq 1/2[1+C^{(1)}]$, and the local interference is the main interference contribution to $R$. On the contrary, when $b\ne b^\prime$, $R\simeq 1/2[1+(2/3g)C^{(2)}]$ and the nonlocal interference is the main interference contribution to $R$. Therefore, $C^{(1)}$ is a ``short-range" interference effect unlike $C^{(2)}$, which is spatially long-range. These properties are reminiscent of classical optics in disordered media, where local interferences explain the granularity of speckle patterns, whereas nonlocal interferences are responsible for long-range correlations in the speckle pattern \cite{Berkovitz94}.
%, defined as
%\begin{equation}
%C^{(2)}\left(x,y\right)=\dfrac{x}{\sqrt{\pi}}\displaystyle\int_0^\infty du\dfrac{3}{\sqrt{2u}\pi}
%\frac{\sinh(\sqrt{2u}\pi)-\sin(\sqrt{2u}\pi)}{\cosh(\sqrt{2u}\pi)-\cos(\sqrt{2u}\pi)}
%\exp(-u^2x^2/4)
%\cos\left(x y u\right).
%\end{equation}

%and
%\begin{equation}
%C^{(1)}(x,y)=\dfrac{x}{\sqrt{\pi}}\displaystyle\int_0^\infty du
%\frac{(\sqrt{2u}\pi)^2}{\cosh(\sqrt{2u}\pi)-\cos(\sqrt{2u}\pi)}
%\exp(-u^2x^2/4)
%\cos\left(x y u\right).
%\end{equation}

\section{Interference, spectral information and Thouless time}
\label{ISI}

The analysis of the previous section reveals that interference, \emph{i.e.}, terms proportional to $C^{(1)}$ and $C^{(2)}$ in Eq. (\ref{Main_result}), depend on the ratio $t_\eta/t_D$. Therefore, they contain information about the spectral properties of the pair, through the characteristic time $t_\eta=t_c/\sqrt{1-\eta}$, and information about the dynamics of the disordered medium, through the Thouless time $t_D$. In this section, we analyze more closely the dependence of $C^{(1)}$ and $C^{(2)}$ on this ratio.
\begin{figure}[h]
\includegraphics[width=10cm]{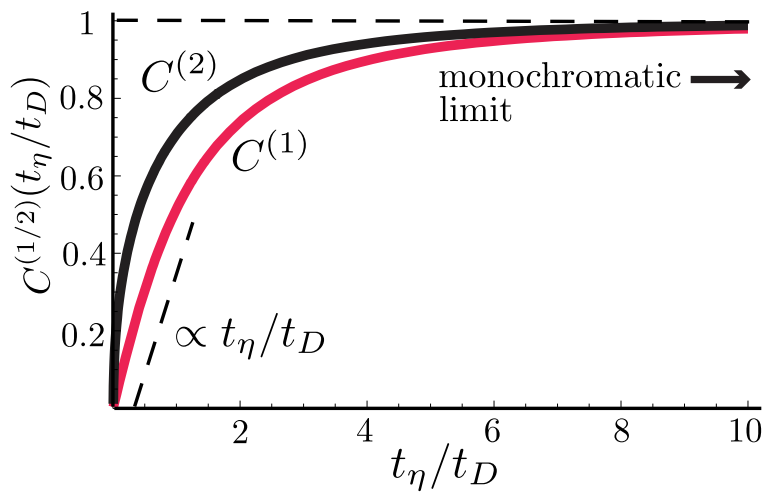}
\caption{\label{Mean_square_Size} 
(color online). Interference contributions $C^{(1)}$ and $C^{(2)}$ to the coincidence counting rate $R$ [see Eq. (\ref{Main_result})], plotted as a function of $t_\eta/t_D$. Both saturate at unity for $t_\eta/t_D\gg1$.
}  
\end{figure}
As shown in Fig. \ref{Mean_square_Size}, both $C^{(1)}$ and $C^{(2)}$ increase with $t_\eta/t_D$, before they saturate at the asymptotic value $C^{(1)}\simeq C^{(2)}\simeq 1$, for $t_\eta\gg t_D$. In this limit, the disordered medium is insensitive to the spectral properties of the photon pair, and the coincidence rate reaches a constant value. This is the monochromatic limit, in which propagation of the pair can be treated within a single-mode approach \cite{Lodahl05, Ott2010}. According to the definition of $t_\eta$, the increase of $C^{(1)}$ and $C^{(2)}$ can be induced in two different manners: by an increase of the coherence time $t_c$ of individual photons, or of the entanglement parameter $\eta$. While the first case simply means that interference between multiply scattered photons can only occur if the individual photon wave-packets overlap (\emph{i.e.}, if $t_c$ is not too small), the second is less obvious. 
\begin{figure}[h]
\includegraphics[width=9cm]{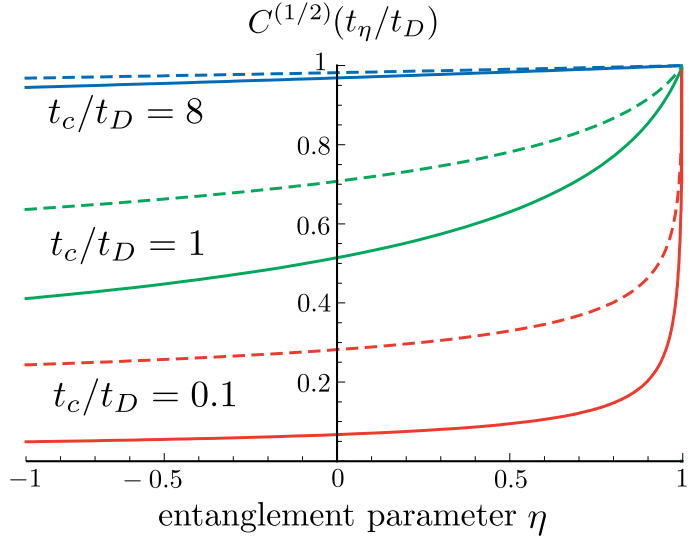}
\caption{\label{C1C2_eta} 
(color online). Functions $C^{(1)}$ (solid curves) and $C^{(2)}$ (dashed curves) plotted as a function of the entanglement parameter $\eta$, for $t_c/t_D=0.1$ (red curves), $t_c/t_D=1$ (green curves) and $t_c/t_D=8$ (blue curves). All curves saturate at unity for strongly correlated photons ($\eta\rightarrow1$).}  
\end{figure}
To clarify the role of entanglement in the interference terms in Eq. (\ref{Main_result}), we show in Fig. \ref{C1C2_eta} $C^{(1)}$ and $C^{(2)}$ as a function of the parameter $\eta$, for three different values of $t_c/t_D$. All curves increase with $\eta$ and saturate at unity when $\eta\rightarrow1$. This can be understood as follows: when $\eta<0$, a given frequency component of one photon tends to be entangled with a \emph{different} frequency component of the other (quantum \emph{anticorrelation}), and therefore does not interfere with it, leading to a reduction of the interference signal after averaging over different realizations of the disorder. Conversely, when $\eta>0$, a given frequency component of the first photon tends to be entangled with the corresponding \emph{same} frequency component of the other (quantum \emph{correlation}), and therefore interferes constructively with it, leading to an enhancement of the interference signal after averaging. From Fig. \ref{C1C2_eta}, we note, however, that the behavior of anticorrelated and correlated photons is not symmetric. In particular, in the limit $\eta\rightarrow1$ of perfectly correlated photons, interference is fully constructive, and photons behave like monochromatic ones, even if the coherence time $t_c$ is small \cite{Campos90}.

\section{Coincidence counting rate with delayed photons}
\label{IDP}

\begin{figure}[h]
\includegraphics[width=11cm]{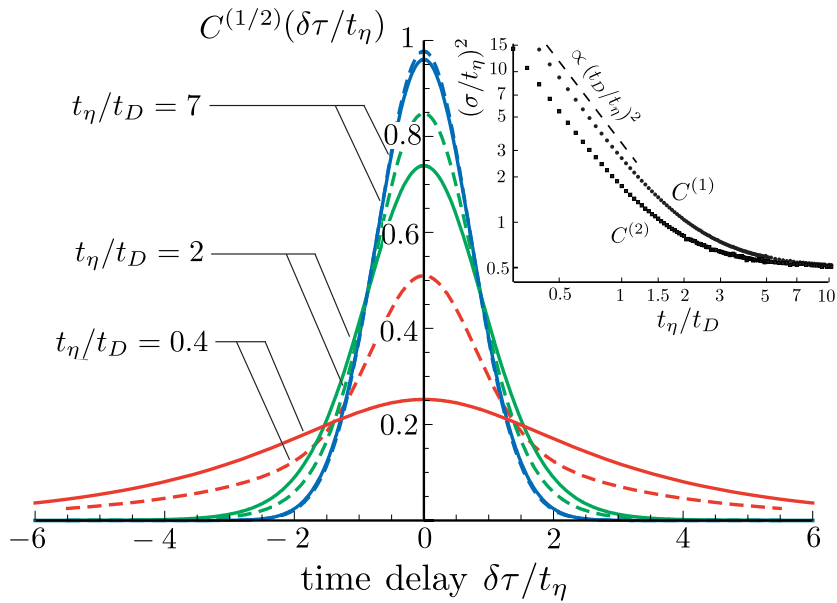}
\caption{\label{photon_pair} 
(color online). Main plot: functions $C^{(1)}$ (solid curves) and $C^{(2)}$ (dashed curves) plotted as a function of the time delay $\delta\tau/t_\eta$, for $t_\eta/t_D=0.4$ (red curves), $t_\eta/t_D=2$ (green curves) and $t_\eta/t_D=7$ (blue curves). Inset: mean square size $(\sigma/t_\eta)^2$ of the functions $C^{(1)}(\delta\tau/t_\eta)$ and $C^{(2)}(\delta\tau/t_\eta)$, as a function of $t_\eta/t_D$ ($\sigma^2$ is defined in the text). For $t_\eta/t_D\gg1$, $\sigma\sim t_\eta$, whereas for $t_\eta/t_D\ll1$, $\sigma\sim t_D$.
}  
\end{figure}

We have shown in the previous section that interference contributions to the coincidence counting rate (\ref{Main_result}) carry information (encoded in the ratio $t_\eta/t_D$) on the entanglement properties of the photon pair and on the dynamics of the disordered medium. We now propose an experimental scheme giving individual access to these two characteristic times. For this purpose, we assume that one of the photons arrives with a small delay $\delta\tau$ with respect to the other (see Fig. \ref{model}). For $\delta\tau\ne0$, the coincidence counting rate (\ref{Main_result}) becomes
\begin{equation}
\label{Main_result2}
R=\dfrac{\overline{\langle{I_bI_{b^\prime}}\rangle}}{\overline{\langle I_b\rangle}\times\overline{\langle I_{b^\prime}\rangle}}=
\dfrac{1}{2}\left\{
1+\dfrac{2}{3g}C^{(2)}\left(\dfrac{t_ \eta}{t_D},\dfrac{\delta\tau}{t_\eta}\right)
+\delta_{bb^\prime}
\left[
C^{(1)}\left(\dfrac{t_ \eta}{t_D},\dfrac{\delta\tau}{t_\eta}\right)+
\dfrac{2}{3g}C^{(2)}\left(\dfrac{t_ \eta}{t_D},\dfrac{\delta\tau}{t_\eta}\right)
\right]
\right\},
\end{equation}
\emph{i.e.}, all interference contributions acquire a dependence on $\delta\tau$. This dependence originates from the fact that the occurence of interference is only possible if the time intervals during which the two photons propagate in the medium overlap. $C^{(1)}$ and $C^{(2)}$ are shown in the main plot of Fig. \ref{photon_pair}, as a function of $\delta\tau/t_\eta$, for several values of $t_\eta/t_D$. We recall that $C^{(1)}$ is the dominant interference contribution to $R$ when $b=b^\prime$ (since $R\simeq1/2[1+C^{(1)}]$), and $C^{(2)}$ when $b\ne b^\prime$ (since $R\simeq1/2[1+(2/3g)C^{(2)}]$).
Two different regimes can be identified from Fig. \ref{photon_pair}: first, when $t_\eta\gg t_D$, $C^{(1)}\simeq C^{(2)}\simeq\exp[-(\delta\tau/t_\eta)^2]$ are peaked functions of $\delta\tau$, and the mean square size $\sigma^2=\int_{-\infty}^\infty  \delta\tau^2C^{(1/2)}(\delta\tau)d(\delta\tau)/\int_{-\infty}^\infty C^{(1/2)}(\delta\tau)d(\delta\tau)\simeq t_\eta^2/2$, as obvious from the asymptotic behavior in the inset of Fig. \ref{photon_pair}. In this limit, measurement of $R$ for different time delays makes the joint coherence time $t_\eta$ of the photon pair accessible. In the opposite limit $t_\eta\ll t_D$, $C^{(1)}$ and $C^{(2)}$ are broad functions of $\delta\tau$, with a reduced visibility. The inset of Fig. \ref{photon_pair} clearly shows that $\sigma\propto t_D$ for both $C^{(1)}(\delta\tau)$ and $C^{(2)}(\delta\tau)$ in this limit. The coincidence counting rate then gives access to the Thouless time, and thus to the diffusion coefficient.

In the diffusive regime considered here, nonlocal interference [which accounts for the $C^{(2)}$ terms in the coincidence counting rate (\ref{Main_result})] is of weak magnitude, and is, therefore, more difficult to measure than local interference. When $b\ne b^\prime$ however, $R\simeq 1/2[1+(2/3g)C^{(2)}]$ [see Eq. (\ref{Main_result2})], and the observation of nonlocal interference may be greatly facilitated by taking advantage of its dependence on $\delta\tau$, which allows us to distinguish it from the constant background of $1/2$. An interesting problem concerns the regime where $g\sim1$, \emph{i.e.}, when Anderson localization sets in. It was shown recently that nonlocal interferences persist in this regime \cite{Ott2010}, where they are of the order of unity and acquire a dependence on the ratio of the sample size $L$ and on the localization length. However, in the context of the present work, their dependence on the spectral properties of the photon pair and on the dynamical properties of the medium is less clear, in particular since the Thouless time is no longer a relevant quantity when $g\sim1$. 

\section{Summary and concluding remarks}

We have calculated the coincidence counting rate that results from the propagation of a pair of entangled photons through a disordered medium, and we have analyzed its dependence on the fourth-order interference that occurs between multiple scattering photon trajectories. We have shown that these interferences convey information on the spectral entanglement of the pair and on the Thouless time of the disordered medium. Finally, for the purpose of measuring this information, we have proposed an experimental scheme in which the photons of the pair are incident on the disordered medium with a time delay. In classical optics, the measurement of the Thouless time usually requires a dynamical experiment, such as time-resolved transmission of  short pulses \cite{Storzer06, Skipetrov06} or spectral correlation mesurements \cite{Albada91}. In the method we propose, no dynamics are involved, only the counting of photon coincidences in the framework of a quantum optics experiment.

Let us conclude with a discussion of the relevance of these results for state of the art experiments. Recently, multiple scattering experiments \cite{Smolka09, Muskens08} were performed on disordered media of titania powders with thicknesses from 5 to 20 $\mu$m, and a mean free path $\ell\simeq 0.9$ $\mu$m. For an average refractive index $n\simeq1.34$ \cite{Muskens08}, this implies Thouless times $t_D=(c/n)\ell/3$ ranging from $3.8\times10^{-14}s$ to $6.0\times10^{-13}$s, with $c$ the speed of light in vacuum. Thus, photons with subpicosecond or subnanosecond coherence time will allow us to explore both scenarios outlined above, $t_\eta\ll t_D$ and $t_\eta\gg t_D$, and to quantify photon entanglement as well as the transport characteristics of the medium.

\section*{Acknowledgments}

We thank V. Shatokhin and S. E. Skipetrov for useful discussions. N. C. acknowledges financial support by the Alexander von Humboldt foundation.


\begin{thebibliography}{99}
\bibitem{Footnote}
The Thouless time is sometimes introduced as $t_D=L^2/D$. In this paper we prefer the definition $t_D=L^2/(\pi^2D)$, which is usually considered in time-resolved experiments in classical optics ($t_D$ then corresponds to the decay rate of the time-dependent transmission of a short pulse across a disordered medium) \cite{Storzer06, Skipetrov06}.
\bibitem{Footnote2}
Four additionnal diagrams also contribute to the four terms in Eqs. (\ref{Main_result}). They are obtained from the diagrams in Fig. \ref{Diagrams} by interchanging $a$ and $a^\prime$.
\end{thebibliography}
\end{document}